\documentclass[showpacs,10pt,twocolumn,prb]{revtex4-1}
\usepackage{amsmath}
\usepackage{amssymb}
\usepackage{graphics}
\usepackage{epsfig}
\usepackage{color}

\begin{document}


\title{Anisotropy in BaFe$_{2}$Se$_{3}$ single crystals with double chains of FeSe tetrahedra}
\author{Hechang Lei,$^{1}$ Hyejin Ryu,$^{1,2}$ Anatoly I. Frenkel,$^{3}$ and C. Petrovic$^{1,2}$}
\affiliation{$^{1}$Condensed Matter Physics and Materials Science Department, Brookhaven National Laboratory, Upton, NY 11973, USA
\\$^{2}$Department of Physics and Astronomy, State University of New York at Stony Brook, Stony Brook, NY 11794, USA
\\$^{3}$Physics Department, Yeshiva University, 245 Lexington Avenue, New York, NY 10016, USA}

\date{\today}

\begin{abstract}
We studied the anisotropy in physical properties of Ba$_{1.00(4)}$Fe$_{1.9(1)}$Se$_{3.1(1)}$ single crystals. BaFe$_{2}$Se$_{3}$ is a semiconductor below 300 K. Magnetization measurements show that there is a crossover from a short-range antiferromangetic (AFM) correlations at room temperature to a long-range AFM order with $T_{N}$ = 255 K. The anisotropy of magnetization is consistent with the previous neutron results. This crossover is supported by the heat capacity measurement where the phase-transition peak is absent at 255 K. The superconducting transition at about 10 K is likely due to the small amount of $\beta$-FeSe impurities.
\end{abstract}

\maketitle

\section{Introduction}

Since the discovery of iron-based superconductors,\cite{Kamihara} iron
pnictide and chalcogenide materials have stimulated great interest. All
iron-based superconductors show some structural similarity. From the
initially discovered LaFeAsO$_{1-x}$F$_{x}$ to very recently reported K$_{x}$%
Fe$_{2-y}$Se$_{2}$ two-dimensional (2D) FePn or FeCh (Pn = pnictogens, Ch =
chalcogens) tetrahedron layers are the common structural ingredient probably
related to high temperature superconductivity.\cite{Kamihara}$^{-}$\cite{Guo}
However, the mechanism of superconductivity and its relation to the crystal
structure motifs in these systems is still under debate. In order to fully
understand the nature of the superconductivity, study of materials
containing similar building blocks related to these systems is of
significant interest.

On the other hand, the dimensionality of magnetic interactions and
electronic transport is another important factor which will influence Fermi
surface topology and magnetic ground state. For example, CaFe$_{4}$As$_{3}$
is a compound which is closely related to iron-based superconductors but has
a different dimensionality and spatial arrangement of similar FePn(Ch) local
structural units. It contains an open three-dimensional (3D) channellike
network of shared FeAs tetrahedra with Ca atoms in the channels that run
along the b axis of the orthorhombic cell.\cite{Todorov} CaFe$_{4}$As$_{3}$
exhibits Fermi-liquid behavior with enhanced electron-electron correlations
and antiferromagnetic (AFM) long range order without superconductivity above
1.8 K.\cite{Todorov}$^{,}$\cite{Zhao}Another structurally related material
is BaFe$_{2}$Se$_{3}$ which contains one-dimensional (1D) double chains of
edge shared Fe-Se tetrahedra along the b-axis.\cite{Hong} Studies of
magnetic properties indicate that there is a crossover from short range AFM
correlation to long range AFM order at around 250 K. However, the existence
of a superconducting transition at the low temperature is still
controversial.\cite{Krzton-Maziopa}$^{,}$\cite{Caron}

In this work, we report the detailed characterization of anisotropic
physical properties and local crystal structure of Ba$_{1.00(4)}$Fe$%
_{1.9(1)} $Se$_{3.1(1)}$ single crystals. We show that this material is a
semiconductor with long-range antiferromagnetic (AFM) order below $T_{N}$ =
255 K.

\section{Experiment}

Single crystals of BaFe$_{2}$Se$_{3}$ were grown by self-flux method with
nominal composition Ba:Fe:Se = 1:2:3. Ba pieces, Fe powder and Se shot were
mixed and put into the carbon crucible, then sealed into the quartz tube
with partial pressure of argon.\ The quartz tube was annealed at 1150 $%
{{}^\circ}%
C$ for 24 h for homogenization, and then the ampoule was slowly cooled down
to 750 $%
{{}^\circ}%
C$ with about 6 $%
{{}^\circ}%
C$/hour. Finally, the furnace was shut down and the ampoule was cooled down
to room temperature naturally. Single crystals with typical size 5$\times $2$%
\times $1 mm$^{3}$ can be grown. The powder X-ray diffraction (XRD) spectra
were taken with Cu K$_{\alpha }$ radiation $\lambda $ = 1.5418 \AA\ using a
Rigaku miniflex x-ray machine. The orientation of crystal is determined
using Bruker SMART APEX II single crystal x-ray diffractometer. The average
stoichiometry was determined by examination of multiple points using an
energy-dispersive x-ray spectroscopy (EDX) in a JEOL JSM-6500 scanning
electron microscope. The X-ray absorption spectra of the Fe and Se $K$-edges
were taken in transmission mode on powder samples of BaFe$_{2}$Se$_{3}$ at
the X18A beamline of the National Synchrotron Light Source. Standard
procedure was used to extract the X-ray absorption near edge structure
(XANES) and extended x-ray absorption fine-structure (EXAFS) from the
absorption spectrum.\cite{Prins} Standard Fe and Se metal foils and FeO, Fe$%
_{2}$O$_{3}$ and Fe$_{3}$O$_{4}$ oxide powders were used for energy
calibration and also for evaluating the valence states of Fe and Se ions.
Electrical transport measurements were performed using a four-probe
configuration. Thin Pt wires were attached to electrical contacts made of
silver paste. Electrical transport, heat capacity, and magnetization
measurements were carried out in Quantum Design PPMS-9 and MPMS-XL5.

\section{Results and Discussion}

Powder XRD result and structural refinements of BaFe$_{2}$Se$_{3}$\ using
Rietica\cite{Hunter} (Fig. 1(a)) indicate that all reflections can be
indexed in the Pnma space group. The refined lattice parameters are a =
11.940(2) \AA , b=5.444(1) \AA , and c = 9.174(1) \AA , consistent with the
values reported in literature.\cite{Krzton-Maziopa}$^{,}$\cite{Caron} The
crystal structure of BaFe$_{2}$Se$_{3}$ (Fig. 1(b)) can be depicted as
alternate stacking of Fe-Se layers and Ba cations along the crystallographic
a axis, similar to K$_{x}$Fe$_{2-y}$Se$_{2}$.\cite{Guo} However, in the
Fe-Se plane, Fe-Se tetrahedra do not connect each other along two planar
directions to form a two-dimensional (2D) infinite layer. Instead, they are
broken along the c-axis periodically and only one dimensional (1D) double
chains of edge shared Fe-Se tetrahedra propagate along the b-axis.. The
double-chains are separated by Ba$^{2+}$ ions and are slightly tilted off
the bc-plane. The tilting directions are opposite between the two
neighboring layers as shown in Fig. 1(d). The shape of BaFe$_{2}$Se$_{3}$
crystal is long plate like. The XRD pattern of a BaFe$_{2}$Se$_{3}$ crystal
(Fig. 1(c)) reveals that the crystal surface is normal to the a axis with
the plate-shaped surface parallel to the bc-plane. Moreover, the b axis is
along the long direction of crystal in the bc-plane. The EDX spectrum of a
single crystal confirms the presence of Ba, Fe and Se. The average atomic
ratios determined from EDX are Ba:Fe:Se = 1.00(4):1.9(1):3.1(1), close to
the expected stoichiometric BaFe$_{2}$Se$_{3}$ ratio. Hence, there are no
vacancies on either Ba or Fe sites, which is different from K$_{x}$Fe$_{2-y}$%
Se$_{2}$.\cite{Guo}$^{,}$\cite{Lei HC}

\begin{figure}[tbp]
\centerline{\includegraphics[scale=0.45]{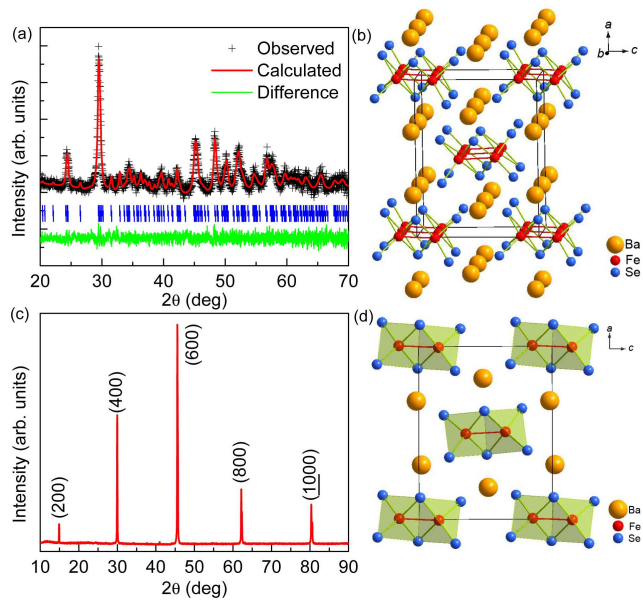}} \vspace*{-0.3cm}
\caption{(a) Powder XRD patterns of BaFe$_{2}$Se$_{3}$. (b) Crystal
structure of BaFe$_{2}$Se$_{3}$. The big orange, small red and medium blue
balls represent Ba, Fe and Se ions. (c) Single crystal XRD of BaFe$_{2}$Se$%
_{3}$. (d) The arrangement of double chains of edge-sharing Fe-Se tetrahedra
along b axis.}
\end{figure}

The features of Fe $K$-edge XANES spectrum of BaFe$_{2}$Se$_{3}$ (Fig. 2(a))
are similar to FeSe$_{x}$.\cite{Chen CL} The pre-peak A ($\sim $ 7112.5 eV)
is due to either dipole or quadrupole 1s $\longrightarrow $ 3d transitions.
The shoulder B with maximum absorption jump at $\sim $ 7118.4 eV can be
ascribed to the 1s $\longrightarrow $ 4p transition. When compared to the
spectrum of FeO, the shoulder B is located at the lower energy side,
suggesting that the valence of Fe is slightly smaller than 2+. The valence
of Fe is evaluated by linear interpolation of the energy of BaFe$_{2}$Se$%
_{3} $ with those of standards (Fe foil (metallic Fe), FeO (Fe$^{2+}$) and Fe%
$_{2} $O$_{3}$ (Fe$^{3+}$)). From the peak of first-derivative spectra
corresponding to the maximum absorption jump we obtain $\sim $ 1.87+ for Fe
valence, which is close to the value for FeSe$_{x}$.\cite{Chen CL} On the
other hand, there are two features for Se $K$-edge XANES spectrum of BaFe$%
_{2}$Se$_{3}$ (Fig. 2b). The feature C is mainly due to the 1s $%
\longrightarrow $ 4p dipole transition and the broad hump D should be a
multiple scattering of the photoelectron with the near neighbors.\cite%
{Joseph} By extrapolating the maximum absorption jumps of Se ($\sim $
12659.2 eV) and SeO$_{2}$ ($\sim $ 12662 eV),\cite{Chen CL} we estimate $%
\sim $ -1.98 for the Se valence, consistent with the formal value of Se ion
(2-). For the Fe site, the nearest neighbors are four Se atoms (1$\times $%
Se(I), 2$\times $Se(II) and 1$\times $Se(III)) with four different distances
close to $\sim $ 2.418 \AA\ and the next nearest neighbors are three Fe
atoms with almost the same distance ($\sim $ 2.723 \AA ).\cite{Hong} From
the joint analysis of Fe and Se edges EXAFS data using a single bond
distance for Fe-Se and Fe-Fe, and by fitting the k range 3.2-18.4 \AA $^{-1}
$ for Fe $K$-edge and 2-14 \AA $^{-1}$ for Se $K$-edge (main panels and
insets of Fig. 2(c) and (d)), the fitted average Fe-Se and Fe-Fe bond
lengths are $d_{Fe-Se}$ = 2.428(5) \AA\ and $d_{Fe-Fe}$ = 2.71(5) \AA . This
is consistent with the reported values.\cite{Hong}
\begin{figure}[tbp]
\centerline{\includegraphics[scale=0.45]{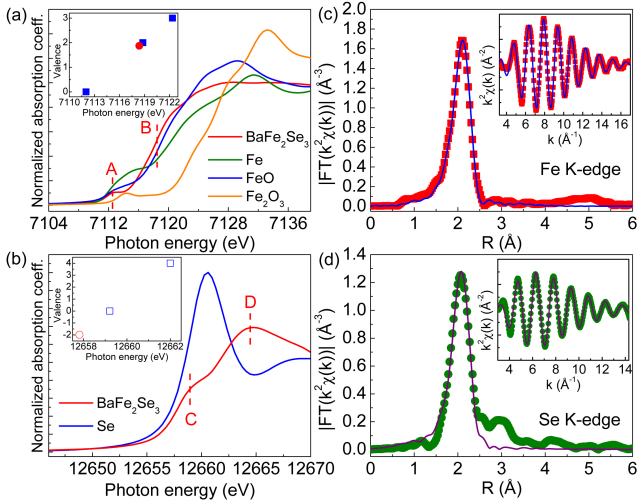}} \vspace*{-0.3cm}
\caption{(a) Fe \textit{K}-edge XANES spectra of BaFe$_{2}$Se$_{3}$, Fe,
FeO, and Fe$_{2}$O$_{3}$ measured at 300 K. The inset shows the valence
states of Fe calculated from the first derivative of the XANES spectra. FT
magnitudes of the EXAFS oscillations (symbols) for Fe K-edge (c) and Se
K-edge (d). The model fits are shown as solid lines. The FTs are not
corrected for the phase shifts and represent raw experimental data. Insets
of (c) and (d) filtered EXAFS (symbols) with k-space model fits (solid
line). }
\end{figure}

Temperature dependence of the resistivity $\rho (T)$ of BaFe$_{2}$Se$_{3}$
crystal for $\mu _{0}H$ = 0 and 9 T indicates that this material is a
semiconductor in measured temperature region. (Fig. 3). The room-temperature
value $\rho (300K)$ is about 17 $\Omega \cdot $cm, which is much larger than
in BaFe$_{2}$S$_{3}$ ($\sim $ 0.35 $\Omega \cdot $cm).\cite{Gonen} Fits of
the $\rho (T)$ at high temperature using the thermal activation model $\rho
=\rho _{0}\exp (E_{a}/k_{B}T)$, where $\rho _{0}$ is a prefactor, $E_{a}$ is
thermal activated energy and $k_{B}$ is Boltzmann's constant, gives $E_{a}$
= 0.178(1) eV in the temperature range above 170 K (inset in Fig. 3), which
is also much larger than in BaFe$_{2}$S$_{3}$.\cite{Gonen}$^{,}$\cite{Reiff}
The larger room-temperature resistivity and $E_{a}$ can be ascribed to the
increase of structural distortion which might localize the carriers and
increase the band gap when compared to BaFe$_{2}$S$_{3}$.\cite{Hong} For BaFe%
$_{2}$S$_{3}$, bond lengths are almost identical for all four Fe-S bonds, in
contrast to very different bond lengths for BaFe$_{2}$Se$_{3}$. Moreover,
the Fe-Fe bond distances in a single chain are identical for BaFe$_{2}$S$%
_{3} $ whereas they are different for BaFe$_{2}$Se$_{3}$, where the longer
and shorter bonds alternately connect along the b axis. All structural
features imply that in the former compound the coordination polyhedron is
more symmetric than in the latter. It should be noted that the shorter Fe-Fe
bond in BaFe$_{2}$S$_{3}$ may also have some contribution to higher
conductivity. On the other hand, when compared to BaFe$_{2}$S$_{3}$ which
exhibits negative magnetoresistance (MR) below 25 K,\cite{Gonen} there is no
obvious MR in BaFe$_{2}$Se$_{3}$ as shown in Fig. 3.
\begin{figure}[tbp]
\centerline{\includegraphics[scale=0.4]{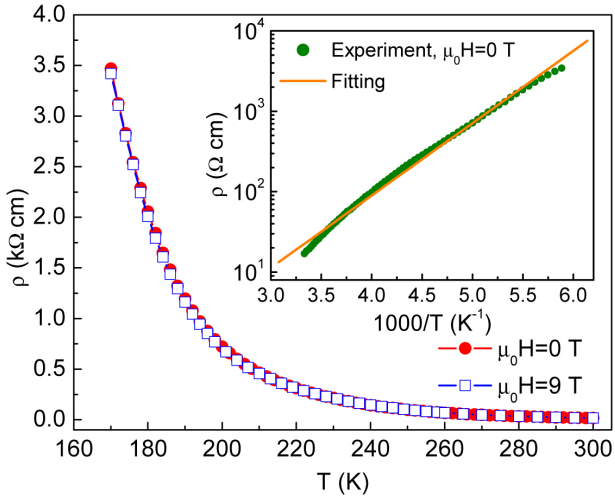}} \vspace*{-0.3cm}
\caption{Temperature dependence of the resistivity $\protect\rho (T)$ of the
BaFe$_{2}$Se$_{3}$ crystal with $\protect\mu _{0}H$ = 0 (closed red circle)
and 9 T (open blue square, H$\Vert $c). Inset: Fitting result of $\protect%
\rho (T)$ at zero field using thermal activation model where the red line is
the fitting curve.}
\end{figure}

\begin{figure}[tbp]
\centerline{\includegraphics[scale=0.43]{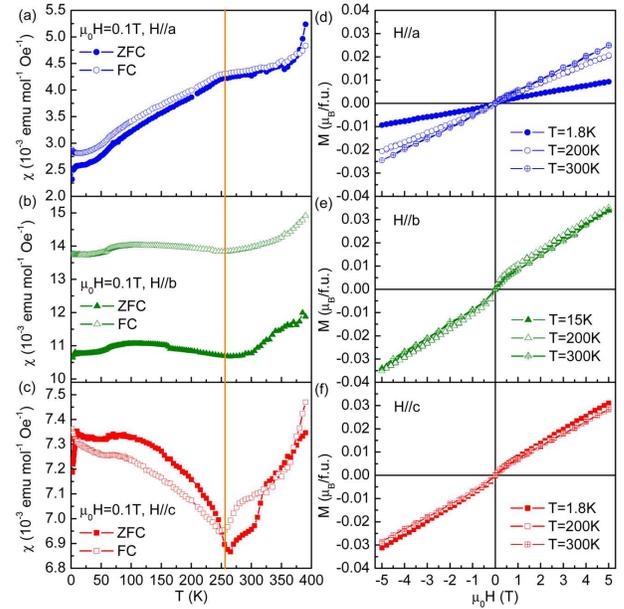}} \vspace*{-0.3cm}
\caption{(a) Temperature dependence of DC magnetic susceptibility $\protect%
\chi $(T) below 300 K under ZFC and FC modes with the applied field $\protect%
\mu _{0}H$ = 0.1 T along (a) a, (b) b and (c) c axis. Isothermal
magnetization hysteresis loops M(H) for (d) H$\Vert $a, (e) H$\Vert $b and
(f) H$\Vert $c at various temperatures.}
\end{figure}

Temperature dependence of dc magnetic susceptibility $\chi (T)$ of BaFe$_{2}$%
Se$_{3}$ single crystal (Fig. 4(a), (b) and (c)) can not be fitted using
Curie-Weiss law up to 390 K. It decreases with decreasing temperature for
all field directions (H$\Vert $a,b, and c). It suggests that there is AFM
interaction up to 390 K, in agreement with neutron results that found short
range correlations (SRC).\cite{Krzton-Maziopa}$^{,}$\cite{Caron} Another
characteristic in $\chi (T)$ curves is a transition appearing at about 255 K
for all three field directions, which is ascribed to a crossover from
short-range AFM correlation to long-range AFM order.\cite{Krzton-Maziopa}$%
^{,}$\cite{Caron} It should be noted that the absolute $\chi _{a}(T)$ is
much smaller than $\chi _{b}(T)$ and $\chi _{c}(T)$. Nevertheless, similar
to H$\parallel $b, a Curie-like upturn in susceptibility is seen below 255 K
for H$\parallel $c, while a faster drop is seen below 255 K for H$\parallel $%
a. It suggests that the easy-axis of magnetization direction is $a$ axis.
According to mean-field theory for collinear antiferromagnet, magnetic
susceptibility along the easy-axis direction goes to zero for T $\rightarrow
$ 0 whereas perpendicular to the easy-axis magnetization direction is nearly
constant below $T_{N}$. This is consistent with the neutron results.\cite%
{Krzton-Maziopa}$^{,}$\cite{Caron} The small positive value of $\chi _{a}(T$
$\rightarrow $ 0$)$ is probably due to the presence of a van Vleck
paramagnetism. A large diversity of susceptibilities between H$\parallel $a,
H$\parallel $b and H$\parallel $c persisting up to 390 K suggests that there
may be an intriguing anisotropy of SRC in the system. On the other hand,
when the magnetic field was applied along the $c$ axis, there is no obvious
hysteresis between ZFC and FC measurements. The lack of hysteresis suggests
that the ferromagnetic component of magnetic interactions between the Fe
chains along the $c$ axis is very weak. This can be ascribed to the large
Fe-Fe (or ferromagnetic Fe4 plaquette) interchain distance ($%
d_{Fe-Fe}^{interchain}$ $\sim $ 6.441 \AA ).\cite{Hong} In contrast, for H$%
\parallel $b, the ZFC and FC measurements exhibit significant hysteresis due
to the fact of short Fe-Fe (or ferromagnetic Fe4 plaquette) intrachain
distance ($d_{Fe-Fe}^{intrachain}$ $\sim $ 2.720 \AA\ and 2.727 \AA ),\cite%
{Hong} resulting in strong magnetic interaction. The hysteresis between the
ZFC and FC curves extends up to 390 K, suggesting that this intrachain
interaction exists at temperatures above the long range order.

Isothermal M(H) for H$\parallel $a, H$\parallel $b and H$\parallel $c (Fig.
4(d), (e) and (f)) at various temperature show that there is no hysteresis
for all three field directions which confirm the AFM order in the system and
exclude the ferromagnetic impurities. For H$\parallel $a, the slopes of M(H)
increase with increasing temperature but for other two directions, M(H)
curves are almost unchanged, consistent with the M(T) results. Moreover, the
slopes of M(H) are H$\parallel $b $>$ H$\parallel $c $>$ H$\parallel $a for
all of measured temperatures, corresponding to the increase of paramagnetic
moment when magnetic field is rotated from a towards c and then towards the
b axis as shown Fig. 4(a), (b) and (c).

\begin{figure}[tbp]
\centerline{\includegraphics[scale=0.4]{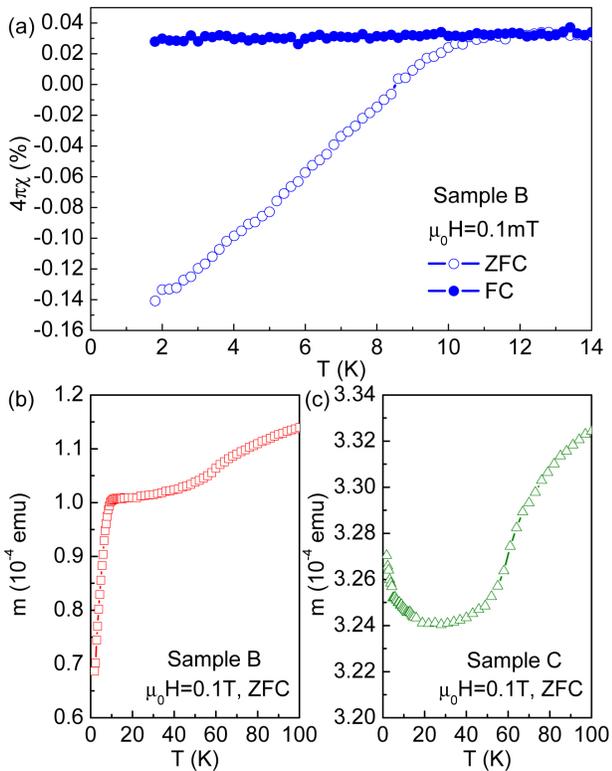}} \vspace*{-0.3cm}
\caption{(a) $\protect\chi $(T) of sample B below 15 K under ZFC and FC
modes with $\protect\mu _{0}H$ = 0.1 mT. $\protect\chi $(T) below 100 K
under ZFC mode with $\protect\mu _{0}H$ = 0.1 T for (b) sample B and (c)
sample C.}
\end{figure}

On the other hand, there is a small drop at about 10 K. This is also observed in previous report and might correspond to the superconducting transition.\cite{Krzton-Maziopa} In order to clarify whether this transition is extrinsic or intrinsic, we measured other two samples from the same batch. As shown in Fig. 5 (a) and (b), sample B exhibits similar drop at about 10 K for $\mu _{0}$H = 0.1 mT and 0.1 T, but the superconducting volume fraction is only about 0.15\% at 1.8 K for $\mu _{0}$H = 0.1 mT. On the other hand, sample C does not show any drop below 15 K for $\mu _{0}$H = 0.1 T. These results imply that this superconducting transition should be extrinsic and coming from residual $\beta$-FeSe in the melt. Moreover, the transition temperature ($\sim $ 10 K) is close to the $T_{c}$ of $\beta$-FeSe ($\sim $ 8.5 K and up to $\sim $ 36.7 K at 8.9 GPa).\cite{Medvedev} Thus, a small drop at about 10 K might be ascribed to the residual amount of $\beta$-FeSe impurities.

\begin{figure}[tbp]
\centerline{\includegraphics[scale=0.4]{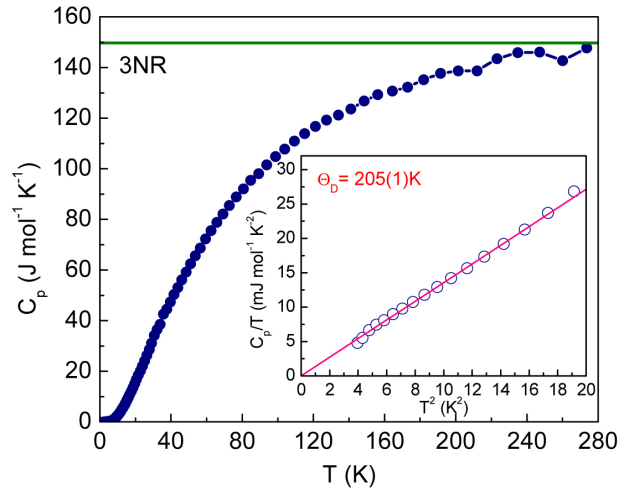}} \vspace*{-0.3cm}
\caption{Temperature dependence of heat capacity for BaFe$_{2}$Se$_{3}$
crystal. The green solid line represents the classical value according to
Dulong-Petit law at high temperature. Inset: the low-temperature
specific-heat data in the plot of C$_{p}$/T vs T$^{2}$. The red solid line
is the fitting curves using formula C$_{p}$/T =$\protect\beta $T$^{2}$.}
\end{figure}

The heat capacity $C_{p}$ of BaFe$_{2}$Se$_{3}$ crystal approaches the
classic value of 3NR at 300 K, where N is the atomic number in the chemical
formula (N = 6) and R is the gas constant (R = 8.314 J mol$^{-1}$ K$^{-1}$),
consistent with the Dulong-Petit law (Fig. 6). At the low temperature, $C_{p}(T)$
curve can be fitted solely by a cubic term $\beta T^{3}$ (inset of Fig. 6).
From the fitted value of $\beta $ = 1.357(6) mJ mol$^{-1}$ K$^{-4}$, the
Debye temperature is estimated to be $\Theta _{D}$ = 205(1) K using the
formula $\Theta _{D}$ = $(12\pi ^{4}NR/5\beta )^{1/3}$. This is slightly
smaller than $\Theta _{D}$ of K$_{x}$Fe$_{2-y}$Se$_{2}$, which might be due
to the larger atomic mass of Ba than K.\cite{Zeng B} On the other hand,
there is no $\lambda $-type anomaly for BaFe$_{2}$Se$_{3}$ at the
temperature of magnetic transitions ($T_{N}$ = 255 K). This can be ascribed
to the overwhelming release of magnetic entropy due to existence of SRC
above the long range order at T$_{N}$.\cite{Caron} This is also consistent
with the absence of Curie-Weiss law in M(T) (Fig. 4(a), (b) and (c)).

\section{Conclusion}

In summary, we studied the physical properties of Ba$_{1.00(4)}$Fe$_{1.9(1)}$%
Se$_{3.1(1)}$ single crystals with 1D double chains of edge shared Fe-Se
tetrahedra parallel to the b-axis. Composition analysis indicates that all
crystallographic sites are fully occupied. XANES result shows that the
valence of Fe is about 1.87+. Taken together with transport, magnetic and
thermodynamic properties, this indicates that the BaFe$_{2}$Se$_{3}$ is a
semiconductor with a short-range AFM correlation at the room temperature and
a long-range AFM order below 255 K. The anisotropy of magnetization is consistent with magnetic structure determined from neutron results previously. This is rather similar to parent
materials of Fe based superconductors, however the absence of appreciable
conductivity coincides with the absence of connected infinite 2D Fe-Se
planar tetrahedra.

\section{Acknowledgement}

We are grateful to Kefeng Wang for helpful discussions. We thank John Warren
for help with scanning electron microscopy measurements and Qi Wang for help
with XAFS measurements. Work at Brookhaven is supported by the U.S. DOE
under Contract No. DE-AC02-98CH10886 and in part by the Center for Emergent
Superconductivity, an Energy Frontier Research Center funded by the U.S.
DOE, Office for Basic Energy Science. A.I.F. acknowledges support by U.S.
Department of Energy Grant DE-FG02-03ER15476. Beamline X18A at the NSLS is
supported in part by the U.S. Department of Energy Grant No
DE-FG02-05ER15688.


\begin{thebibliography}{99}
\bibitem{Kamihara} Y. Kamihara, T. Watanabe, M. Hirano, and H. Hosono, J.
Am. Chem. Soc. \textbf{130}, 3296 (2008).

\bibitem{Rotter} M. Rotter, M. Tegel, and D. Johrendt, Phys. Rev. Lett.
\textbf{101}, 107006 (2008).

\bibitem{Wang XC} X. C. Wang, Q. Q. Liu, Y. X. Lv, W. B. Gao, L. X. Yang, R.
C. Yu, F. Y. Li ,and C. Q. Jin, Solid State Commun. \textbf{148}, 538 (2008).

\bibitem{Hsu FC} F. C. Hsu, J. Y. Luo, K. W. Yeh, T. K. Chen, T. W. Huang,
P. M. Wu, Y. C. Lee, Y. L. Huang, Y. Y. Chu, D. C. Yan, and M. K. Wu, Proc.
Natl. Acad. Sci. USA \textbf{105}, 14262 (2008).

\bibitem{Guo} J. Guo, S. Jin, G. Wang, S. Wang, K. Zhu, T. Zhou, M. He, and
X. Chen, Phys. Rev. B \textbf{82}, 180520(R) (2010).

\bibitem{Todorov} I. Todorov, D. Y. Chung, C. D. Malliakas, Q. Li, T. Bakas,
A. Douvalis, G. Trimarchi, K. Gray, J. F. Mitchell, A. J. Freeman, and M. G.
Kanatzidis, J. Am. Chem. Soc. \textbf{131}, 5405 (2009).

\bibitem{Zhao} L. L. Zhao, T. H. Yi, J. C. Fettinger, S. M. Kauzlarich, and
E. Morosan, Phys. Rev. B \textbf{80}, 020404(R) (2009).

\bibitem{Hong} H. Y. Hong and H. Steinfink, J. Solid State Chem. \textbf{5},
93 (1972).

\bibitem{Krzton-Maziopa} A. Krzton-Maziopa, E. Pomjakushina, V. Pomjakushin,
D. Sheptyakov, D. Chernyshov, V. Svitlyk, and K. Conder, J. Phys.: Condens.
Matter \textbf{23}, 402201 (2011).

\bibitem{Caron} J. M. Caron, J. R. Neilson, D. C. Miller, A. Llobet, and T.
M. McQueen, Phys. Rev. B \textbf{84}, 180409(R) (2011).

\bibitem{Prins} R. Prins and D. Koningsberger, X-ray Absorption: Principles,
Applications, Techniques of EXAFS, SEXAFS, XANES, Wiley, New York (1988).

\bibitem{Hunter} Hunter B. (1998) "Rietica - A visual Rietveld program",
International Union of Crystallography Commission on Powder Diffraction
Newsletter No. 20, (Summer) http://www.rietica.org

\bibitem{Lei HC} H. C. Lei and C. Petrovic, Phys. Rev. B \textbf{83}, 184504
(2011).

\bibitem{Chen CL} C. L. Chen, S. M. Rao, C. L. Dong, J. L. Chen, T. W.
Huang, B. H. Mok, M. C. Ling, W. C. Wang, C. L. Chang, T. S. Chan, J. F.
Lee, J.-H. Guo, and M. K. Wu, EPL, 93 47003 (2011).

\bibitem{Joseph} B. Joseph, A. Iadecola, L. Simonelli, Y. Mizuguchi, Y.
Takano, T. Mizokawa, and N. L. Saini, J. Phys.: Condens. Matter \textbf{22},
485702 (2010).

\bibitem{Gonen} Z. S. G\"{o}nen, P. Fournier, V. Smolyaninova, R. Greene, F.
M. Araujo-Moreira, and B. Eichhorn, Chem. Mater. \textbf{12}, 3331 (2000).

\bibitem{Reiff} W. M. Reiff, I. E. Grey, A. Fan, Z. Eliezer, and H.
Steinfink, J. Solid State Chem. \textbf{13}, 32 (1975).

\bibitem{Medvedev} S. Medvedev, T. M. McQueen, I. A. Troyan, T. Palasyuk, M.
I. Eremets, R. J. Cava, S. Naghavi, F. Casper, V. Ksenofontov, G.Wortmann,
and C. Felser, Nature Mater. \textbf{8} 630 (2009).

\bibitem{Zeng B} B. Zeng, B. Shen, G. F. Chen, J. B. He, D. M. Wang, C. H.
Li, and H. H. Wen, Phys. Rev. B \textbf{83}, 144511 (2011).
\end{thebibliography}
\end{document}